\documentclass{IEEEoj}
\usepackage{cite}
\usepackage{amsmath,amssymb,amsfonts}
\usepackage{algorithmic}
\usepackage{graphicx,color}
\usepackage{textcomp}
\def\BibTeX{{\rm B\kern-.05em{\sc i\kern-.025em b}\kern-.08em
    T\kern-.1667em\lower.7ex\hbox{E}\kern-.125emX}}
\AtBeginDocument{\definecolor{ojcolor}{cmyk}{0.93,0.59,0.15,0.02}}
\def\OJlogo{\vspace{-4pt}\hskip-4pt\includegraphics[height=18pt]{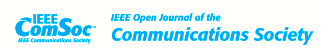}}

\newcommand{\rb}{{\textbf{r}}}

\usepackage{algorithm}

\begin{document}
\receiveddate{XX Month, XXXX}
\reviseddate{XX Month, XXXX}
\accepteddate{XX Month, XXXX}
\publisheddate{XX Month, XXXX}
\currentdate{11 January, 2024}
\doiinfo{OJCOMS.2024.011100}

\title{3D 8-Ary Noise Modulation Using Bayesian- and Kurtosis-based Detectors}

\author{Hadi Zayyani\IEEEauthorrefmark{1}\IEEEmembership{(Member, IEEE)}, Felipe A. P. de Figueiredo\IEEEauthorrefmark{2}\IEEEmembership{(Senior Member, IEEE)}, Mohammad Salman\IEEEauthorrefmark{3}\IEEEmembership{(Senior Member, IEEE)}, And Rausley A. A. de Souza\IEEEauthorrefmark{2}\IEEEmembership{(Senior Member, IEEE)}}
\affil{Department of Electrical and Computer Engineering, Qom University of Technology (QUT), Qom, Iran}
\affil{National Institute of Telecommunications (Inatel), Santa Rita do Sapucaí, Brazil}
\affil{College of Engineering and Technology, American University of the Middle East, Egaila, 54200, Kuwait}
\corresp{CORRESPONDING AUTHOR: Hadi Zayyani (e-mail: zayyani@qut.ac.ir).}
\authornote{This work has been partially funded by the xGMobile Project (XGM-AFCCT-2024-9-1-1) with resources from EMBRAPII/MCTI (Grant 052/2023 PPI IoT/Manufatura 4.0), by CNPq (302085/2025-4, 306199/2025-4), and FAPEMIG (APQ-03162-24).}
\markboth{Preparation of Papers for IEEE OPEN JOURNALS}{Author \textit{et al.}}

\begin{abstract}
This paper presents a novel three-dimensional (3D) 8-ary noise modulation scheme that introduces a new dimension: the mixture probability of a Mixture of Gaussian (MoG) distribution. This proposed approach utilizes the dimensions of mean and variance, in addition to the new probability dimension. Within this framework, each transmitted symbol carries three bits, each corresponding to a distinct sub-channel.
For detection, a combination of specialized detectors is employed: a simple threshold-based detector for the first sub-channel bit (modulated by the mean), a Maximum-Likelihood (ML) detector for the second sub-channel bit (modulated by the variance), a Kurtosis-based, Jarque-Bera (JB) test, and Bayesian Hypothesis (BHT)-based detectors for the third bit (modulated by the MoG probability). The Kurtosis- and JB-based detectors specifically distinguish between Gaussian (or near-Gaussian) and non-Gaussian MoG distributions by leveraging higher-order statistical measures.
The Bit Error Probabilities (BEPs) are derived for the threshold-, Kurtosis-, and BHT-based detectors. The optimum threshold for the Kurtosis-based detector is also derived in a tractable manner. Simulation results demonstrate that a comparably low BEP is achieved for the third sub-channel bit relative to existing two-dimensional (2D) schemes. Simultaneously, the proposed scheme increases the data rate by a factor of 1.5 and 3 compared to the Generalized Quadratic noise modulator and the classical binary KLJN noise modulator, respectively. Furthermore, the Kurtosis-based detector offers a low-complexity solution, achieving an acceptable BEP of approximately 0.06.
\end{abstract}


\maketitle

\section{INTRODUCTION}
\label{sec:Intro}

\IEEEPARstart{N}{oise} modulation is a breakthrough modulation approach for the newborn noise communication framework. In the noise communication framework, noisy waveforms are used instead of carrier waveforms. This leads to zero- or low-power communication schemes.
Although the concept of exploiting the thermal noise of resistors with high and low variances was first proposed by Kish over 20 years ago \cite{Kish05}, its widespread adoption only followed a pivotal communication-theoretic reframing of the idea by Basar in 2023 \cite{Basar23}.
After the pioneering work of \cite{Kish05}, the same author proposed a Kirchhoff-Law-Johnson Noise (KLJN) secure key exchange scheme, in which the laws of physics provide the basis for unconditionally secure communication \cite{Kish06}. In addition, Kish et al. proposed totally secure classical networks with multi-point telecloning (teleportation) of classical bits through loops with Johnson-like noise \cite{Kish06_1}.

Following Kish's foundational work, more advanced studies on the KLJN noise communication scheme emerged, exploring various aspects of its core principle. This body of research incorporated practical considerations for high-frequency operation \cite{Ming08}, investigated the impact of wire resistance on noise voltage and current \cite{Kish10}, and made initial attempts to calculate the system's BEP \cite{Saez13}. Subsequent efforts focused on reducing the BEP further by leveraging both voltage and current noises \cite{Saez13_1} and developing more advanced BEP calculation methods \cite{Smul14}, among other related topics \cite{Ging14}-\cite{Kape22}.

As previously noted, the field was significantly advanced by Basar's 2023 communication-theoretic framework \cite{Basar23}. This work inspired a new series of developments, including: a proposed noise modulator that feeds noise samples directly into an antenna system \cite{Basar24}; an analysis of thermal noise modulation through optimal detection and performance metrics \cite{Alshaw24}; and an on-off noise modulator design \cite{Anjos25}.

Recent innovations continue to expand the paradigm. \cite{Tasci25} proposed a Flip-KLJN secure communication scheme where a pre-agreed trigger flips the bit mapping during exchange. \cite{Salem25} introduced an asymmetric KLJN-based modulation using a four-resistor structure to improve Bit Error Rate (BER) without extra samples per bit, presenting a viable solution for ultra-low-power IoT security. \cite{Yapici25} developed a novel joint energy harvesting and communication scheme for IoT devices based on noise modulation. Furthermore, \cite{ZayyArxiv25} proposed a Generalized Quadratic Noise Modulator (GQNM) utilizing non-Gaussian distributions to double the data rate, and \cite{ZayyArxivRH25} suggested a resistor-hopping system to increase the data rate based on the number of chips per bit duration.

In this paper, we propose a three-dimensional, 8-ary noise modulator scheme to extend the order of noise modulation beyond the two-dimensional framework presented in \cite{ZayyArxiv25}. The proposed scheme utilizes two established dimensions mean and variance and introduces a novel third dimension: the mixture probability of a Mixture of Gaussian (MoG) distribution.

This new dimension modulates the distribution's Gaussianity. A low mixture probability (e.g., for a two-component MoG) results in a nearly Gaussian distribution, while a high probability yields a distinctly non-Gaussian MoG. This property enables the encoding of a third bit per symbol to select between a quasi-Gaussian and a non-Gaussian MoG distribution.

We investigate different detectors for this modulator and calculate the BEP for the sub-channels where a closed-form or tractable analysis is feasible. Specifically, a threshold-based detector is used for the mean dimension bit, a Maximum Likelihood (ML) detector for the variance dimension bit, and a Kurtosis-based detector for the newly introduced bit governing the MoG distribution. We derive the BEP in a closed-form or tractable manner for the threshold-based and Kurtosis-based detectors. Furthermore, the optimal threshold for the Kurtosis-based detector is calculated theoretically.

Simulation results demonstrate successful detection across all three sub-bit channels. This scheme increases the data rate by a factor of 1.5 compared to the Quadratic noise modulator \cite{ZayyArxiv25} and by a factor of 3 compared to the classical KLJN noise modulator.

The paper is organized as follows. After the literature survey in the introduction, Section~\ref{sec:ProblemForm} discusses the system model and preliminaries. In Section~\ref{sec: prop}, the proposed 3-dimensional 8-ary noise modulation framework is presented and then developed in detail in Section~\ref{sec: MoG}. Moreover, the BEP analysis is presented in Section~\ref{sec: BEP}. Simulation results are presented in Section~\ref{sec: Simulation}, while conclusions are drawn in Section~\ref{sec: con}.

\section{System Model and Preliminaries}\label{sec:ProblemForm}

In this section, we briefly discuss the basic concepts of the classic KLJN noise modulator and the GQNM modulator that were presented in \cite{Basar24} and \cite{ZayyArxiv25}, respectively.

The work in \cite{Basar24} describes a modulator where, based on the conveyed bit, low- and high-variance noise samples are generated using the thermal noise from low and high resistances. These noise samples are then fed directly to an antenna. At the receiver, baseband processing identifies whether the variance is low or high to detect the information bits. This KLJN noise modulator is fundamentally a binary scheme operating on a single variance dimension.

In \cite{ZayyArxiv25}, a GQNM is proposed. This scheme introduces a second dimension bias voltage or mean to augment the variance dimension, creating a two-dimensional quadratic noise modulator. Within this framework, two bits per symbol are used to adjust the mean and variance independently. This GQNM scheme also utilizes non-Gaussian noise distributions, such as a MoG or a Laplacian distribution, which is reported to provide unequal BEP across the two sub-channels and can reduce the BEP in the variance sub-channel. Consequently, the quadratic modulator achieves both a higher data rate and a lower overall BEP in its sub-channels compared to its binary predecessor.

To further increase the data rate and develop higher-order modulation schemes, we propose a three-dimensional, 8-ary noise modulator in this paper, which will be discussed in the next section.

\begin{figure}
\begin{center}
\includegraphics[scale=0.34]{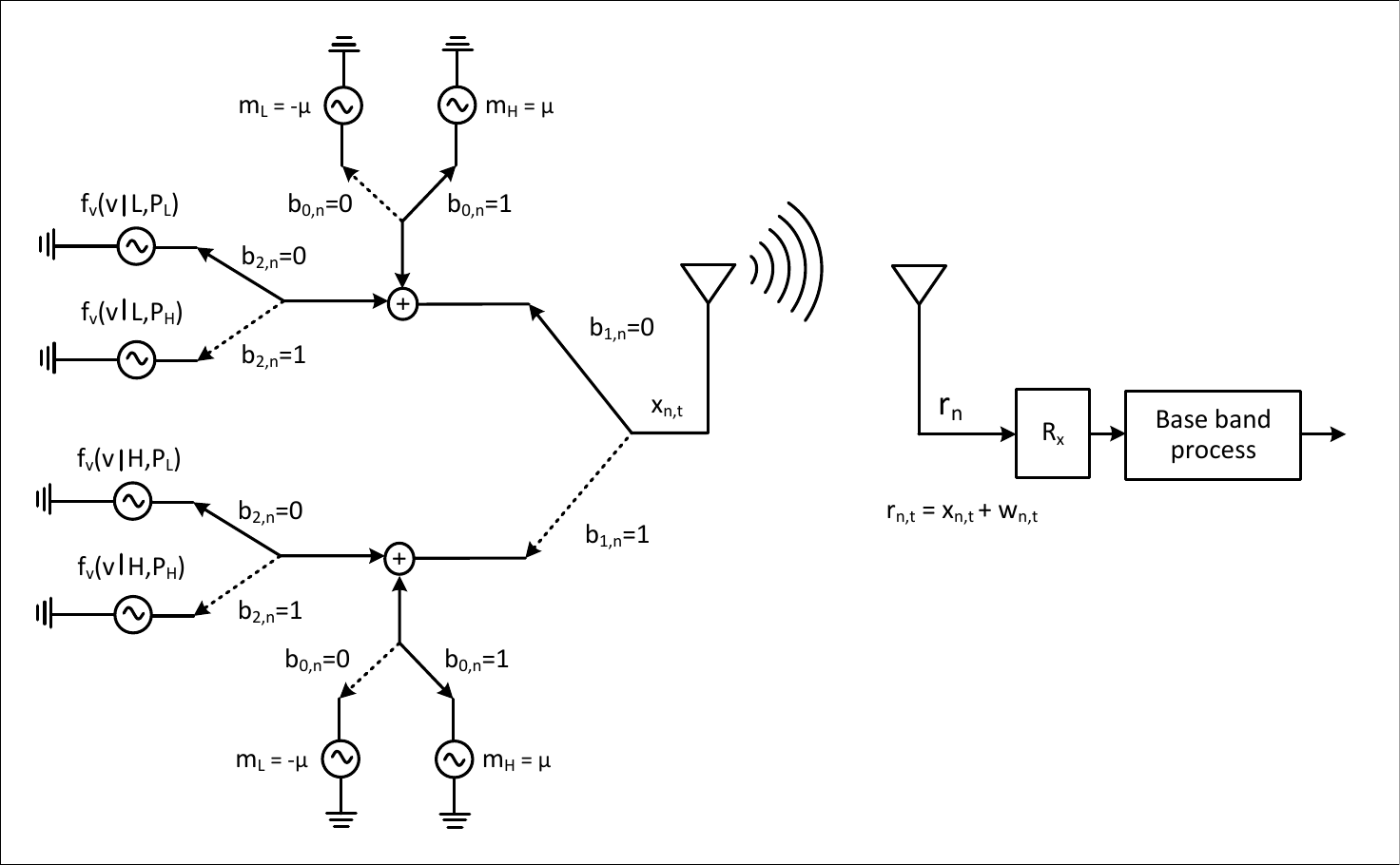}
\end{center}
\vspace{-0.5 cm}
\caption{Block diagram of the 3D 8-ary noise modulation scheme.}
\label{fig1}
\end{figure}

\begin{figure}
\begin{center}
\includegraphics[scale=0.55]{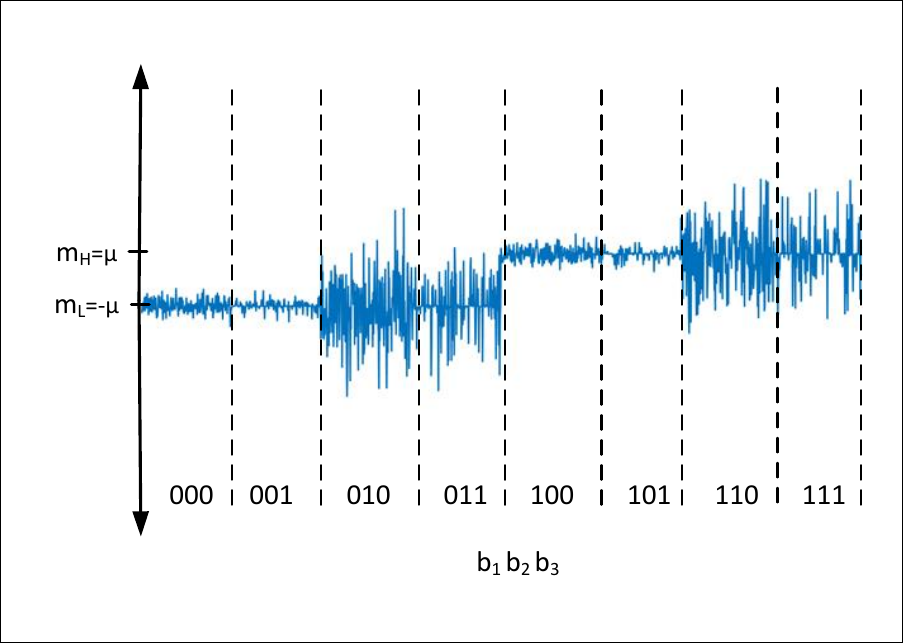}
\end{center}
\vspace{-0.5 cm}
\caption{Examples of waveforms of the 3D 8-ary noise modulation scheme.}
\label{fig2}
\end{figure}

\section{3D 8-ary mixture-based noise modulation}
\label{sec: prop}

In this section, we present the proposed 3D 8-ary Mixture-Based Noise Modulation (3D-8-ary-MBNM) scheme, as illustrated in Fig.~1. The scheme encodes three information bits, denoted as $b_0$, $b_1$, and $b_2$.

The first bit, $b_0$, selects the voltage bias to modulate the mean dimension. For improved separability, the mean values are set to $m_L=-\mu$ and $m_H=\mu$. Specifically, if $b_0=0$, the mean is $m_L=-\mu$; if $b_0=1$, the mean is $m_H=+\mu$.

The second bit, $b_1$, selects the mixture distribution to modulate the variance dimension. If $b_1=0$, a mixture distribution with low variance, $f_v(v|L)$, is selected. Conversely, if $b_1=1$, a mixture distribution with high variance, $f_v(v|H)$, is chosen.

For the novel third dimension, we introduce the mixture probability. We assume a mixture distribution of the form
\begin{align}
p(v)=pf_v(v|L)+(1-p)f_v(v|H),
\end{align}
where $p$ is the mixture probability, $f_v(v|L)$ is a Probability Density Function (PDF), such as Gaussian or Laplacian, for the low variance case, and $f_v(v|H)$ is the corresponding PDF for the high variance case. When $p = p_L$ is a low probability near zero, the resulting mixture distribution approximates a pure, non-mixture distribution. Conversely, when $p = p_H$ is a high probability (typically around 0.5\footnote{A value near 1 would again result in an approximately pure PDF.}), the distribution becomes a true mixture.

The third bit, $b_2$, leverages this property: a value of $b_2=0$ selects a pure non-Gaussian distribution (e.g., by setting $p_L=0$), while $b_2=1$ selects a mixture distribution with the high probability $p_H$. The incorporation of this third dimension enables an 8-ary modulation scheme. Typical waveforms for all eight symbols of this Mixture-Based Noise Modulation (MBNM) scheme are illustrated in Fig.~2.

In this study, we focus on a Gaussian Mixture model. We develop the corresponding modulators and investigate several detectors, including a simple threshold-based detector, an ML detector, and a Kurtosis-based detector. For the two simpler detectors, we derive the BEPs mathematically, resulting in closed-form expressions or tractable formulations. These derivations will be presented in the following section.



\section{MBNM modulator with Gaussian mixture distributions}
\label{sec: MoG}

\subsection{General description}
\label{sec: genMoG}

The MBNM modulator utilizes MoG-distributed noise samples. The MoG distribution has the form $(1-p)\mathcal{N}(0,\sigma^2_{\text{off}}) + p\mathcal{N}(0,\sigma^2_{\text{on}})$, where $\mathcal{N}(0,\sigma^2) = \frac{1}{\sigma\sqrt{2\pi}}\exp\left(-\frac{v^2}{2\sigma^2}\right)$ is the Gaussian distribution and $\sigma_{\text{off}} \ll \sigma_{\text{on}}$. By excluding the mean values $m_H = \mu$ or $m_L = -\mu$, we obtain four distinct zero-mean Gaussian mixture distributions\footnote{When $p=0$ or $p=1$, the distribution reverts to a pure Gaussian.}
\begin{align}
f_v(v|b_1b_2=00)=(1-p_L)N(0,\sigma^2_{00})+p_LN(0,\sigma^2_{10}),
\end{align}
\begin{align}
f_v(v|b_1b_2=01)=(1-p_H)N(0,\sigma^2_{00})+p_HN(0,\sigma^2_{10}),
\end{align}
\begin{align}
f_v(v|b_1b_2=10)=(1-p_L)N(0,\sigma^2_{01})+p_LN(0,\sigma^2_{11}),
\end{align}
\begin{align}
f_v(v|b_1b_2=11)=(1-p_H)N(0,\sigma^2_{01})+p_HN(0,\sigma^2_{11}),
\end{align}
where $\sigma^2_{00}=\sigma^2_{\text{off},L}$ is the variance of the first (smaller) component in the low-variance MoG case ($b_1=0$), $\sigma^2_{10}=\sigma^2_{\text{on},L}$ is the variance of the second (larger) component in the low-variance MoG case, $\sigma^2_{01}=\sigma^2_{\text{off},H}$ is the variance of the first (smaller) component in the high-variance MoG case ($b_1=1$), and $\sigma^2_{11}=\sigma^2_{\text{on},H}$ is the variance of the second (larger) component in the high-variance MoG case.

The variances are scaled such that $\sigma^2_{10}=\alpha\sigma^2_{00} \gg \sigma^2_{00}$, $\sigma^2_{11}=\alpha\sigma^2_{01} \gg \sigma^2_{01}$, $\sigma^2_{01}=\beta\sigma^2_{00} \gg \sigma^2_{00}$, and $\sigma^2_{11}=\beta\sigma^2_{10} \gg \sigma^2_{10}$. This is achieved by selecting scaling factors $\alpha \gg 1$ and $\beta \gg 1$. Thus, integrating this into a unified model per Fig.~1, the transmitted noise sample $x_{n,t}$—where $n$ is the symbol index and $t \in {1,2,\dots,N}$ is the noise sample index within a symbol of length $N$—follows an MoG distribution given by
\begin{align}
x_{n,t}\sim (1-p_n)N(m_n,\sigma^2_{0,n})+p_nN(m_n,\sigma^2_{1,n}),
\end{align}
where we have
\begin{align}
\label{eq: mn}
m_n=\left\{
      \begin{array}{ll}
        \mu, & b_{0,n}=0, \\
        -\mu, & b_{0,n}=1,
      \end{array}
    \right.
\end{align}
and
\begin{align}
\label{eq: sig0n}
\sigma^2_{0,n}=\left\{
      \begin{array}{ll}
        \sigma^2_{00}=\sigma^2_{off,L}, & b_{1,n}=0, \\
        \sigma^2_{01}=\sigma^2_{off,H}, & b_{1,n}=1,
      \end{array}
    \right.
\end{align}
and
\begin{align}
\label{eq: sig1n}
\sigma^2_{1,n}=\left\{
      \begin{array}{ll}
        \sigma^2_{10}=\sigma^2_{on,L}, & b_{1,n}=0, \\
        \sigma^2_{11}=\sigma^2_{on,H}, & b_{1,n}=1,
      \end{array}
    \right.
\end{align}
and
\begin{align}
p_n=\left\{
      \begin{array}{ll}
        p_L, & b_{2,n}=0, \\
        p_H, & b_{2,n}=1.
      \end{array}
    \right.
\end{align}

In the detector step at the receiver, we utilize the sample mean, $\hat{m}_n$, and the sample power, $\hat{P}_n$. For a large number of samples $N$, and by the Central Limit Theorem (CLT), these estimators are given by
\begin{align}
\label{eq: mhat}
\hat{m}_n=\frac{1}{N}\sum_{t=1}^Nr_{n,t}\sim N(m_n,\sigma^2_{\hat{m}}),
\end{align}
\begin{align}
\label{eq: varhat}
\hat{P}_n=\frac{1}{N}\sum_{t=1}^Nr^2_{n,t}\sim N(m_{p,n},\mathrm{Var}(\hat{P}_n)),
\end{align}
where we have
\begin{align}
\label{eq: vv}
\sigma^2_{\hat{m}}&=\mathrm{Var}(\hat{m})=\frac{1}{N}\mathrm{Var}(r_{n,t})=\frac{1}{N}[\mathrm{Var}(x_{n,t})+\sigma^2_w],\nonumber\\
&=\frac{1}{N}\Big[(1-p_n)\sigma^2_{0,n}+p_n\sigma^2_{1,n}+\sigma^2_w\Big],
\end{align}
where $\mathrm{Var}(x_{n,t})$ is equal to
\begin{align}
\label{eq: varxn}
\mathrm{Var}(x_{n,t})=\left\{
                        \begin{array}{ll}
                          (1-p_L)\sigma^2_{00}+p_L\sigma^2_{10}, & (b_{1,n},b_{2,n})=00 \\
                          (1-p_L)\sigma^2_{01}+p_L\sigma^2_{11}, & (b_{1,n},b_{2,n})=10 \\
                          (1-p_H)\sigma^2_{00}+p_H\sigma^2_{10}, & (b_{1,n},b_{2,n})=01 \\
                          (1-p_H)\sigma^2_{01}+p_H\sigma^2_{11}. & (b_{1,n},b_{2,n})=11
                        \end{array}
                      \right.
\end{align}
and
\begin{align}
\label{eq: mpn}
&m_{p,n}=\mathrm{E}\{\hat{P}_n\}=\mathrm{E}\{r^2_{n,t}\}=\mathrm{E}\{x^2_{n,t}\}+\sigma^2_w\nonumber\\
&=\mathrm{Var}(x_{n,t})+m^2_n+\sigma^2_w=\nonumber\\
&\left\{
                        \begin{array}{ll}
                          V_{00}\triangleq(1-p_L)\sigma^2_{00}+p_L\sigma^2_{10}+\mu^2+\sigma^2_w, & (b_{1,n},b_{2,n})=00 \\
                          V_{10}\triangleq(1-p_L)\sigma^2_{01}+p_L\sigma^2_{11}+\mu^2+\sigma^2_w, & (b_{1,n},b_{2,n})=10 \\
                          V_{01}\triangleq(1-p_H)\sigma^2_{00}+p_H\sigma^2_{10}+\mu^2+\sigma^2_w, & (b_{1,n},b_{2,n})=01 \\
                          V_{11}\triangleq(1-p_H)\sigma^2_{01}+p_H\sigma^2_{11}+\mu^2+\sigma^2_w, & (b_{1,n},b_{2,n})=11
                        \end{array}
                      \right.
\end{align}
where we have $V_{00}<V_{01}<V_{10}<V_{11}$ which is simply hold. The PDF of $\hat{P}_n$ is a mixture of four Gaussians with the means equal to $V_{ij}$ for $0\le i,j\le 1$.
Moreover, the variance of $\hat{P}_n$ which is nominated as $\mathrm{Var}(\hat{P}_n)\triangleq\sigma^2_v$ is equal to
\begin{align}
\label{eq: f1}
\mathrm{Var}(\hat{P}_n)=\sigma^2_v=\frac{1}{N}\mathrm{Var}(r^2_{n,t})=\frac{1}{N}[\mathrm{E}\{r^4_{n,t}\}-\mathrm{E}^2\{r^2_{n,t}\}],
\end{align}
where we have
\begin{align}
\mathrm{E}\{r^2_{n,t}\}=m_{p,n},
\end{align}
where $m_{p,n}$ is given in (\ref{eq: mpn}), and we have
\begin{align}
\mathrm{E}\{r^4_{n,t}\}&=\mathrm{E}\{(x_{n,t}+w_{n,t})^4\}\nonumber\\
&=\mathrm{E}\{x^4_{n,t}\}+6\mathrm{E}\{x^2_{n,t}\}\sigma^2_w+\sigma^2_w,
\end{align}
where some calculations show that we have
\begin{align}
\label{eq: ff}
\mathrm{E}\{x^2_{n,t}\}=\mathrm{Var}\{x_{n,t}\}+\mu^2,
\end{align}
and
\begin{align}
\label{eq: fff}
\mathrm{E}\{x^4_{n,t}\}=\mathrm{E}\{(Y+m_n)^4\}=\mathrm{E}\{Y^4\}+6\mathrm{E}\{Y^2\}\mu^2+\mu^4,
\end{align}
where we have
\begin{align}
\mathrm{E}\{Y^4\}=(1-p_n)\sigma^2_{0,n}+p_n\sigma^2_{1,n},\\
\mathrm{E}\{Y^4\}=3(1-p_n)\sigma^4_{0,n}+3p_n\sigma^4_{1,n},
\end{align}
which is obtained from the second and fourth-order moments of a MoG distribution. However, the variances for the four Gaussian cases of the random variable $\hat{P}_n$ are calculated. We avoid their closed-form expressions due to their complexity and instead denote the variances of the four Gaussian distributions as $\sigma^2_v\Big|_{(b_1,b_2)=ij} = \sigma^2_{v,ij}$ for simplicity in BEP calculations throughout the paper.

\subsection{Detectors}

\subsubsection{Simple detector for detecting $b_{0,n}$}

As from general derivations in subsection~\ref{sec: genMoG}, the statistics of $\hat{m}$ consists of a mixture of 8 Gaussian distributions in which 4 Gaussian distributions are around $m_n=-\mu$ and four others are around $m_n=+\mu$. Hence, the detector for the bit $b_{0,n}$ is simply given as
\begin{align}
\label{eq: detb0}
\hat{b}_{0,n}=\left\{
            \begin{array}{ll}
              0, & \hat{m}<0, \\
              1, & \hat{m}>0.
            \end{array}
          \right.
\end{align}

Moreover, from the discussions in Subsection~\ref{sec: genMoG}, the statistic $\hat{P}_n$ follows a mixture of four Gaussian distributions with means $V{ij}\Big|{(i,j)=(b_1,b_2)}$ and variances $\sigma^2_v\Big|_{(i,j)=(b_1,b_2)}$. To facilitate the derivation of simple threshold-based detectors, we assume these Gaussian distributions are sufficiently separated relative to their standard deviations. Specifically, the following distinguishability conditions are assumed to hold
\begin{align}
\label{eq: conds}
V_{00}+3\sigma_{v,00}\ll V_{10}-3\sigma_{v,10},\\
V_{10}+3\sigma_{v,10}\ll V_{01}-3\sigma_{v,01},\\
V_{01}+3\sigma_{v,01}\ll V_{11}-3\sigma_{v,11}.
\end{align}

By this assumptions, the simple detector for the bits $b_{1,n}$ and $b_{2,n}$ is
\begin{align}
(\hat{b}_{1,n},\hat{b}_{2,n})=\left\{
                                \begin{array}{ll}
                                  00, & \hat{P}_n\le\mathrm{Th}_{v,1} \\
                                  10, &  \mathrm{Th}_{v,1}\le\hat{P}_n\le\mathrm{Th}_{v,2}\\
                                  01, & \mathrm{Th}_{v,2}\le\hat{P}_n\le\mathrm{Th}_{v,2} \\
                                  11, & \hat{P}_n\ge\mathrm{Th}_{v,3},
                                \end{array}
                              \right.
\end{align}
where the simple thresholds are selected as $\mathrm{Th}_{v,1}\triangleq \frac{V_{00}+V_{01}}{2}$, $\mathrm{Th}_{v,2}\triangleq \frac{V_{10}+V_{01}}{2}$, and $\mathrm{Th}_{v,3}\triangleq \frac{V_{10}+V_{11}}{2}$.

\subsubsection{ML detector}

Simulation results indicate that the distinguishability conditions in (\ref{eq: conds}) are not satisfied. Specifically, the mean $V_{00}$ is too close to $V_{01}$ relative to their variances, and similarly, $V_{10}$ is too close to $V_{11}$. Consequently, the simple detectors exhibit poor performance, resulting in a high BEP for the sub-channels $b_{1,n}$ and $b_{2,n}$. An additional factor is the non-Gaussian nature of the MoG distribution, which necessitates metrics beyond simple mean and variance. Therefore, we resort to an optimal detector, namely the ML detector, defined as follows
\begin{align}
\label{eq: LL}
&\mathrm{ML}: \quad \mathrm{Max}_{b_{1,n},b_{2,n}}\quad p(\rb_n|b_{1,n},b_{2,n})=\prod_{t=1}^Tp(r_{n,t}|b_{1,n},b_{2,n})\nonumber\\
&=\prod_{t=1}^N\Big[(1-p_n)\frac{1}{\sigma_{0,n}\sqrt{2\pi}}\exp(-\frac{(r_{n,t}-m_n)}{2\sigma^2_{0,n}})+\nonumber\\
&p_n\frac{1}{\sigma_{1,n}\sqrt{2\pi}}\exp(-\frac{(r_{n,t}-m_n)}{2\sigma^2_{1,n}})\Big],
\end{align}
where $\rb_n=[r_{n,1}, r_{n,2},..., r_{n,N}]^T$ is the vector of received sample noises in a symbol duration. The likelihood in (\ref{eq: LL}) is calculated over all four possible cases of $(b_{1,n},b_{2,n})=00, 10, 01, 11$, and the maximum argument is detected as the detected bits.

Furthermore, simulations of the ML detector reveal that while it can correctly detect $b_{1,n}$, its performance in detecting $b_{2,n}$ is unacceptable. This is due to the non-Gaussian nature of the MoG and the statistical resemblance between the two MoG distributions governed by $p_L$ and $p_H$, which makes distinguishing between them to detect $b_{2,n}$ particularly challenging. Consequently, we employ a Kurtosis-based detector, which is presented next.

\subsubsection{Normality-test detector based on Kurtosis}

In essence, detecting the bit $b_{2,n}$ requires distinguishing between two MoG distributions with mixture probabilities $p_H$ and $p_L$. In the special case of $p_L = 0$ employed in our simulations, this problem simplifies to distinguishing a pure Gaussian distribution from a non-Gaussian MoG distribution with probability $p_H$. This is a problem of normality testing \cite{Agos86}. While a rich literature exists on Gaussianity tests \cite{Agos86}, a detailed review is beyond the scope of this paper. One of the simplest and most efficient methods for this test is to use kurtosis \cite{Derrik98}. Kurtosis, a fourth standardized moment, is defined for a random variable $X$ as
\begin{align}
\label{eq: Kor}
\mathrm{Kurtosis}(X)=\mathrm{E}\{\Big(\frac{X-\mu}{\sigma}\Big)^4\}=\frac{\mu_4}{\mu^2_2},
\end{align}
where $\mu_4 = \mathrm{E}{(X - \mu)^4}$ and $\mu_2 = \mathrm{E}{(X - \mu)^2} = \sigma^2$ is the second moment (variance) of $X$. The underlying principle is to use kurtosis to distinguish between the case where $r_{n,t}$ follows a Gaussian distribution (with kurtosis equal to 3) and the case where it follows a non-Gaussian MoG distribution (with kurtosis typically greater than 3). We therefore propose the following kurtosis-based detector for $b_{2,n}$, to be applied after detecting $b_{0,n}$ using a simple threshold detector and $b_{1,n}$ using an ML detector
\begin{align}
\hat{b}_{2,n}=\left\{
                \begin{array}{ll}
                  1, & \mathrm{Kor}(r_{n,t})>\mathrm{Th}_k \\
                  0, & \mathrm{Otherwise}
                \end{array}
              \right.
\end{align}
where $\mathrm{Kor}(.)$ stands for the kurtosis function which is defined in (\ref{eq: Kor}), and $\mathrm{Th}_k>3$ is a threshold. Some hints on how to optimize the threshold and calculate the BEP of sub-bit channel $b_{2,n}$ are provided in the next part of the discussion about the BEP of sub-bit channels.

\subsubsection{Normality-test detector based on the Jarque-Bera test}

In statistics, the Jarque-Bera (JB) test is a goodness-of-fit test used to assess whether sample data exhibit the skewness and kurtosis of a normal distribution \cite{Jark87}. The test statistic is given by \cite{Hall95}
\begin{align}
\mathrm{JB}=\frac{N}{6}(S^2+\frac{1}{4}(K-3)^2),
\end{align}
where $\mathrm{JB}$ is the JB test statistics, $K=\mathrm{Kur}(X)$ is the Kurtosis, where $X=r_{n,t}$, and $S=\frac{\mu_3}{\sigma^3}$ is the skewness, where $\mu_3=\mathrm{E}\{(X-\mu)^3\}=\sigma^2$. To detect $b_{2,n}$, we compare the JB statistics with a threshold as
\begin{align}
\hat{b}_{2,n}=\left\{
                \begin{array}{ll}
                  1, & \mathrm{JB}(r_{n,t})>\mathrm{Th}_{JB} \\
                  0, & \mathrm{Otherwise}
                \end{array}
              \right.
\end{align}
where $\mathrm{Th}_{JB}$ is the threshold of the JB test. In the simulations, we find an optimum JB threshold.

\subsubsection{Bayesian-Hypothesis Testing Detector}
\label{sec: BHTdet}

In deriving this detector, we assume the bits $b_{0,n}$ and $b_{1,n}$ have already been detected. We employ a Bayesian Hypothesis Testing (BHT) approach for detecting $b_{2,n}$, deciding in favor of hypothesis $H_0: b_{2,n} = 0$ over $H_1: b_{2,n} = 1$ when the following condition holds:
\begin{align}
p(\rb_n|H_0)>p(\rb_n|H_1),
\end{align}
where we have
\begin{align}
p(\rb|H_0)=\prod_{t=1}^N&\Big[\frac{p_L}{\sigma_{0n}\sqrt{2\pi}}\exp(-\frac{(r_{n,t}-m_n)^2}{2\sigma^2_{0n}})+\nonumber\\
&\frac{1-p_L}{\sigma_{0n}\sqrt{2\pi}}\exp(-\frac{(r_{n,t}-m_n)^2}{2\sigma^2_{0n}})\Big],
\end{align}
and
\begin{align}
p(\rb|H_1)=\prod_{t=1}^N&\Big[\frac{p_H}{\sigma_{0n}\sqrt{2\pi}}\exp(-\frac{(r_{n,t}-m_n)^2}{2\sigma^2_{0n}})+\nonumber\\
&\frac{1-p_H}{\sigma_{0n}\sqrt{2\pi}}\exp(-\frac{(r_{n,t}-m_n)^2}{2\sigma^2_{0n}})\Big].
\end{align}

\section{BEP calculation}
\label{sec: BEP}

This section analyzes the Bit Error Probability (BEP) calculation for the proposed 3D-8ary MBNM scheme under Gaussian mixture assumptions. The analysis covers three distinct sub-channels corresponding to bits $b_{0,n}$, $b_{1,n}$, and $b_{2,n}$. We derive the BEP for $b_{0,n}$ using a simple threshold-based detector, and for $b_{2,n}$ using both Kurtosis-based and BHT detectors. The BEP calculation for $b_{1,n}$, which employs an ML detector, presents significant challenges and is left for future work.

\subsection{BEP calculation of sub-bit $b_{0,n}$}

The first sub-bit channel BEP is calculated as
\begin{align}
\label{eq: pb0}
p_{b_{0}}&=\mathrm{p}\{\hat{b}_{0,n}\neq b_{0,n}\}\nonumber\\
&=\frac{1}{2}\mathrm{p}\{\hat{b}_{0,n}\neq b_{0,n}|b_{0,n}=1\}+\frac{1}{2}\mathrm{p}\{\hat{b}_{0,n}\neq b_{0,n}|b_{0,n}=0\}\nonumber\\
&=\frac{1}{2}\mathrm{p}\{\hat{m}_n<0|b_{0,n}=1\}+\frac{1}{2}\mathrm{p}\{\hat{m}_n>0|b_{0,n}=0\}\nonumber\\
&=\mathrm{p}\{\hat{m}_n>0|b_{0,n}=0\},
\end{align}
where $\mathrm{p}\{\hat{m}_n<0|b_{0,n}=1\}=\mathrm{p}\{\hat{m}_n>0|b_{0,n}=0\}$. Due to symmetry, we have
\begin{align}
\label{eq: p14}
&\mathrm{p}\{\hat{m}_n>0|b_{0,n}=0\}=\frac{1}{4}\mathrm{p}\{\hat{m}_n>0|(b_{0,n},b_{1,n},b_{2,n})=000\}\nonumber\\
&+\frac{1}{4}\mathrm{p}\{\hat{m}_n>0|(b_{0,n},b_{1,n},b_{2,n})=001\}\nonumber\\
&+\frac{1}{4}\mathrm{p}\{\hat{m}_n>0|(b_{0,n},b_{1,n},b_{2,n})=010\}\nonumber\\
&+\frac{1}{4}\mathrm{p}\{\hat{m}_n>0|(b_{0,n},b_{1,n},b_{2,n})=011\}.
\end{align}

From (\ref{eq: mhat}), we have
\begin{align}
\label{eq: pmij}
\mathrm{p}\{\hat{m}_n>0|(b_{0,n},b_{1,n},b_{2,n})=0ij\}=\mathrm{Q}\Big(\frac{\mu}{\sigma_{\hat{m}}\Big|_{ij}}\Big),
\end{align}
where $(i,j)=(b_{1,n},b_{2,n})$, and from (\ref{eq: vv}), we have
\begin{align}
&\sigma^2_{\hat{m}}\Big|_{ij}=\frac{1}{N}\Big[(1-p_n)\sigma^2_{0,n}+p_n\sigma^2_{1,n}+\sigma^2_w\Big]\Big|_{(b_{1,n},b_{2,n})=ij}\nonumber\\
&=\left\{
                        \begin{array}{ll}
                          \frac{1}{N}[(1-p_L)\sigma^2_{00}+p_L\sigma^2_{10}+\sigma^2_w], & ij=00 \\
                          \frac{1}{N}[(1-p_L)\sigma^2_{01}+p_L\sigma^2_{11}+\sigma^2_w], & ij=10 \\
                          \frac{1}{N}[(1-p_H)\sigma^2_{00}+p_H\sigma^2_{10}+\sigma^2_w], & ij=01 \\
                          \frac{1}{N}[(1-p_H)\sigma^2_{01}+p_H\sigma^2_{11}+\sigma^2_w. & ij=11
                        \end{array}
                      \right.
\end{align}

Replacing (\ref{eq: pmij}) into (\ref{eq: p14}) and then in (\ref{eq: pb0}), we have a compact formula of BEP for the sub-bit channel of $b_{0,k}$, which is
\begin{align}
p_{b_{0}}=\frac{1}{4}\sum_{i=0}^1\sum_{j=0}^1\mathrm{Q}\Big(\frac{\mu}{\sigma_{\hat{m}}|_{ij}}\Big).
\end{align}

\subsection{BEP of sub-bit channel of $b_{2,n}$}

\subsubsection{BEP of BHT detector}

To calculate the BEP of $b_{2,n}$ based on the BHT detector, which is presented in subsection \ref{sec: BHTdet}, we can write
\begin{align}
\label{eq: BEPBHT}
&\mathrm{p}_{b,2,BHT}=\frac{1}{2}\mathrm{p}\{e|b_{2,n}=0\}+\frac{1}{2}\mathrm{p}\{e|b_{2,n}=1\}\nonumber\\
&=\frac{1}{2}\mathrm{p}\{T_{1,ki}>T_{0,ki}|b_{2,n}=0\}+\frac{1}{2}\mathrm{p}\{T_{0,ki}>T_{1,ki}|b_{2,n}=1\}\nonumber\\
&=\frac{1}{8}\sum_{k=0}^1\sum_{i=0}^1\Big[\mathrm{p}\{T_{1,ki}>T_{0,ki}|b_{2,n}=0\}\Big]\nonumber\\
&+\frac{1}{8}\sum_{k=0}^1\sum_{i=0}^1\Big[\mathrm{p}\{T_{0,ki}>T_{1,ki}|b_{2,n}=1\}\Big],
\end{align}
where $(k,i)=(b_{0,n},b_{1,n})$, and we have
\begin{align}
\label{eq: T0ij}
&T_{0,ki}\triangleq \sum_{t=1}^N\ln \Big[\frac{p_L}{\sigma_{0n}\sqrt{2\pi}}\exp(-\frac{(r_{n,t}-m_n)^2}{2\sigma^2_{0n}})+\nonumber\\
&\frac{1-p_L}{\sigma_{0n}\sqrt{2\pi}}\exp(-\frac{(r_{n,t}-m_n)^2}{2\sigma^2_{0n}})\Big]\sim N(m_{0,ki},\sigma^2_{0,ki}),
\end{align}
and
\begin{align}
\label{eq: T1ij}
&T_{1,ki}\triangleq \sum_{t=1}^N\ln \Big[\frac{p_H}{\sigma_{0n}\sqrt{2\pi}}\exp(-\frac{(r_{n,t}-m_n)^2}{2\sigma^2_{0n}})+\nonumber\\
&\frac{1-p_H}{\sigma_{0n}\sqrt{2\pi}}\exp(-\frac{(r_{n,t}-m_n)^2}{2\sigma^2_{0n}})\Big]\sim N(m_{1,ki},\sigma^2_{1,ki}),
\end{align}
where $T_{0,ki}$ and $T_{1,ki}$ are Gaussian for $N\gg 1$ due to the CLT theorem, even the distributions of the terms like
\begin{align}
\label{eq: Fij}
F_{ki}\triangleq &\ln\Big[\frac{p_L}{\sigma_{0n}\sqrt{2\pi}}\exp(-\frac{(r_{n,t}-m_n)^2}{2\sigma^2_{0n}})+\nonumber\\
&\frac{1-p_L}{\sigma_{0n}\sqrt{2\pi}}\exp(-\frac{(r_{n,t}-m_n)^2}{2\sigma^2_{0n}})\Big]\Big|_{(b_0,b_1)=ki}
\end{align}
and
\begin{align}
\label{eq: Gij}
G_{ki}\triangleq &\Big[\frac{p_H}{\sigma_{0n}\sqrt{2\pi}}\exp(-\frac{(r_{n,t}-m_n)^2}{2\sigma^2_{0n}})+\nonumber\\
&\frac{1-p_H}{\sigma_{0n}\sqrt{2\pi}}\exp(-\frac{(r_{n,t}-m_n)^2}{2\sigma^2_{0n}})\Big]\Big|_{(b_0,b_1)=ki},
\end{align}
are very complicated to compute. We can assume $F_{ki}$ has mean $m_{F,ki}$ and variance $\sigma^2_{F,ki}$, and similarly $G_{ki}$ has mean of mean $m_{G,ki}$ an and variance $\sigma^2_{G,ki}$. Then, we have $m_{0,ki}=Nm_{F,ki}$, $m_{1,ki}=Nm_{G,ki}$, $\sigma^2_{0,ki}=N^2\sigma^2_{F,ki}$, and $\sigma^2_{1,ki}=N^2\sigma^2_{G,ki}$.

If we assume that $Z=T_{0,ki}\sim N(m_Z,\sigma^2_Z)$ and $W=T_{1,ki}\sim N(m_W,\sigma^2_W)$ are two independent normal random variables we have
\begin{align}
\label{eq: pzw}
\mathrm{p}\{Z>W\}=\mathrm{p}\{\tilde{Z}+m_Z>\tilde{W}+m_W\}=\mathrm{p}\{\tilde{Z}>\tilde{W}+\delta\},
\end{align}
where $\delta\triangleq m_W-m_Z$, $\tilde{Z}=Z-m_Z\sim N(0,\sigma^2_Z)$, and $\tilde{W}=W-m_W\sim N(0,\sigma^2_W)$. Then, using (\ref{eq: pzw}), we have
\begin{align}
\label{eq: pzwQ}
&\mathrm{p}\{Z>W\}=\mathrm{p}\{A>B+\delta\}\int_{-\infty}^{+\infty}\int_{\delta+a}^{+\infty}f_a(a)f_b(b)d_ad_b\nonumber\\
&=\int_{-\infty}^{+\infty}f_b(b)(\int_{\delta+b}^{+\infty}f_a(a)d_a)d_b\nonumber\\
&=\int_{-\infty}^{+\infty}f_b(b)\mathrm{Q}\Big(\frac{\delta+b}{\sigma_A}\Big)d_b\nonumber\\
&=\frac{1}{\sigma_B\sqrt{2\pi}}\int_{-\infty}^{+\infty}\exp(-\frac{b^2}{\sigma^2_A})\mathrm{Q}\Big(\frac{\delta+b}{\sigma_A}\Big)d_b,
\end{align}
where $A=\tilde{Z}$ and $B=\tilde{W}$. To calculate the integral in (\ref{eq: pzwQ}), we use the approximation of $\mathrm{Q}(r)\approx \exp(-ar^2-b_r-c)$, where $a=0.3842$, $b=0.7640$, and $c=0.6964$ \cite{Benitez11}. Then, we can write
\begin{align}
\label{eq: qqint}
\mathrm{p}\{Z>W\}=\frac{1}{\sigma_B\sqrt{2\pi}}\int_{-\infty}^{+\infty}\exp(-A_1b^2-B_1b-C_1)db,
\end{align}
where $A_1\triangleq \frac{1}{\sigma^2_B}+\frac{a^2}{\sigma^2_A}$, $B_1\triangleq \frac{2a\delta}{\sigma^2_A}+\frac{b}{\sigma_A}$, and $C_1\triangleq \frac{a\delta^2}{\sigma^2_A}+\frac{b\delta}{\sigma_A}+c$.

Finally, writing $-A_1b^2-B_1b-C_1=-A_1(b-b_w)^2+c_w$, where $b_w\triangleq \frac{B_1}{2A_1}$, and $c_w\triangleq -C_1+2A_1 b^2_w$. Then, from (\ref{eq: qqint}), with some calculations, we find
\begin{align}
\label{eq: pzwfinal}
\mathrm{p}\{Z>W\}=\frac{\exp(c_w)}{\sigma_B}\sqrt{\frac{1}{2A_1}}.
\end{align}

From (\ref{eq: pzwfinal}) and using that in (\ref{eq: BEPBHT}), the calculation of the BEP is tractable, but not in closed form. To put forward a more straightforward formula for the BEP, from (\ref{eq: BEPBHT}) and (\ref{eq: pzwfinal}), we can write
\begin{align}
\label{eq: BEPBHTfinal}
&\mathrm{p}_{b,2,BHT}=\frac{1}{8}\sum_{k=0}^1\sum_{i=0}^1\Big[1-\frac{\exp(c_{w,ki})}{\sigma_{1,ki}}\sqrt{\frac{1}{2A_{1,ki}}}\Big]+\nonumber\\
&+\frac{1}{8}\sum_{k=0}^1\sum_{i=0}^1\Big[\frac{\exp(c_{w,ki})}{\sigma_{1,ki}}\sqrt{\frac{1}{2A_{1,ki}}}\Big],
\end{align}
where we have
\begin{align}
c_{w,ki}=-C_{1,ki}+2A_{1,ki} b^2_{w,ki},
\end{align}
where $C_{1,ki}=\frac{a\delta^2}{\sigma^2_{0,ki}}+\frac{b\delta}{\sigma_{0,ki}}+c$, $A_{1,ki}=\frac{1}{\sigma^2_{1,ki}}+\frac{a^2}{\sigma^2_{0,ki}}$, and $b_{w,ki}=\frac{B_{1,ki}}{2A_{1,ki}}$ where $B_{1,ki}=\frac{2a\delta}{\sigma^2_{0,ki}}+\frac{b}{\sigma_{0,ki}}$. In the aforementioned formulas, we have $\sigma_{0,ki}=N\sigma_{F,ki}$ and $\sigma_{1,ki}=N\sigma_{G,ki}$, where $F_{ki}$ and $G_{ki}$ are defined in (\ref{eq: Fij}) and (\ref{eq: Gij}).

\subsubsection{BEP of Kurtosis-based detector}
\label{sec: BEPKur}

To calculate the BEP of sub-bit channel of $b_{2,n}$, we have
\begin{align}
\label{eq: BEPK}
&\mathrm{p}_{b,2}=\mathrm{p}\{\hat{b}_{2,n}\neq b_{2,n}\}\nonumber\\
&=\frac{1}{2}\mathrm{p}\{error|b_{2,n}=0\}+\frac{1}{2}\mathrm{p}\{error|b_{2,n}=1\}\nonumber\\
&=\frac{1}{2}\mathrm{p}\{\mathrm{\hat{Kur}}_1>\mathrm{Th}_k\}+\frac{1}{2}\mathrm{p}\{\mathrm{\hat{Kur}}_2<\mathrm{Th}_k\},
\end{align}
where $\mathrm{\hat{Kur}}_1=\frac{\hat{\mu}_{4,0}}{\hat{\mu^2}_{2,0}}$ is the calculated kurtosis in the first case where we have $b_{2,n}=0$, and $\mathrm{\hat{Kur}}_1=\frac{\hat{\mu}_{4,1}}{\hat{\mu^2}_{2,1}}$ is the calculated kurtosis in the first case where we have $b_{2,n}=1$. We have
\begin{align}
\hat{\mu}_{4,i}=\frac{1}{N}\sum_{t=1}^N\Big(r_{n,t}-\mathrm{E}(r_{n,t})\Big)^4\sim N(\mu_{4,i},\sigma^2_{4,i}),
\end{align}
and
\begin{align}
\label{eq: mu2i}
\hat{\mu}_{2,i}=\frac{1}{N}\sum_{t=1}^N\Big(r_{n,t}-\mathrm{E}(r_{n,t})\Big)^2\sim N(\mu_{2,i},\sigma^2_{2,i}),
\end{align}
where
\begin{align}
&\mu_{4,i}=\mu_4(r_{n,t})\Big|_{b_{2,n}=i}=\mathrm{E}\{(r_{n,t}-m_n)^4\}\Big|_{b_{2,n}=i}=\nonumber\\
&\mathrm{E}\{\tilde{r}^4_{n,t}\}\Big|_{b_{2,n}=i},
\end{align}
where $\tilde{r}{n,t} \sim p_n \mathcal{N}(0,\sigma^2{0n}+\sigma^2_w) + (1-p_n)\mathcal{N}(0,\sigma^2_{1n}+\sigma^2_w)$, with $p_n = p_L$ for $i = 0$ and $p_n = p_H$ for $i = 1$. Here, $\sigma_{0n}$ and $\sigma_{1n}$ are given by (\ref{eq: sig0n}) and (\ref{eq: sig1n}), respectively, and are known since $b_{1,n}$ is detected prior to detecting $b_{2,n}$. Therefore, we have
\begin{align}
&\mu_{4,i}=3p_n(\sigma^2_{0n}+\sigma^2_w)^2+3(1-p_n)(\sigma^2_{1n}+\sigma^2_w)^2,
\end{align}
where $i=b_{2,n}$. Additionally, from (\ref{eq: mu2i}), and following some simple calculations, we have
\begin{align}
\mu_{2,i}=p_n(\sigma^2_{0n}+\sigma^2_w)+(1-p_n)(\sigma^2_{1n}+\sigma^2_w).
\end{align}
To calculate $\sigma^2_{2,i}$, we have
\begin{align}
&\sigma^2_{2,i}=\mathrm{Var}(\hat{\mu}_{2,i})=\frac{1}{N}\mathrm{Var}(\tilde{r}^2_{n,t})=\nonumber\\
&\frac{1}{N}\Big[p_n\mathrm{Var}(\tilde{r}^2_{n,t})\Big|_0+(1-p_n)\mathrm{Var}(\tilde{r}^2_{n,t})\Big|_1\Big]\nonumber\\
&\frac{2}{N}\Big[p_n\sigma^4_{0n}+(1-p_n)\sigma^4_{1n}\Big],
\end{align}
where in calculations, we use $\mathrm{Var}(n^2)=\mathrm{E}(n^4)-\mathrm{E}^2(n^2)=3\sigma^4_n-\sigma^4_n=2\sigma^4_n$.
To calculate $\sigma^2_{4,i}$, we have
\begin{align}
\sigma^2_{4,i}=\frac{1}{N}\Big[p_n\mathrm{Var}(\tilde{r}^4_{n,t})\Big|_0+(1-p_n)\mathrm{Var}(\tilde{r}^4_{n,t})\Big|_1\Big],
\end{align}
where $\mathrm{Var}(\tilde{r}^4_{n,t})\Big|_0=96\sigma^8_{0n}$ since for a zero=mean Gaussian random variable $n$, we have $\mathrm{Var}(n^4)=\mathrm{E}\{n^8\}-\mathrm{E}^2\{n^4\}=105\sigma^8_n-9\sigma^8_n=96\sigma^8_n$. Moreover, we have $\mathrm{Var}(\tilde{r}^4_{n,t})\Big|_1=96\sigma^8_{1n}$. This way, we have
\begin{align}
\sigma^2_{4,i}=\frac{96}{N}\Big[p+_n\sigma^8_{0n}+(1-p_n)\sigma^8_{1n}\Big].
\end{align}
To put forward the BEP calculations in (\ref{eq: BEPK}), we have
\begin{align}
\mathrm{\hat{Kur}}_i=\frac{\hat{\mu}_{4,i}}{\hat{\mu^2}_{2,i}},
\end{align}
where $\hat{\mu}_{4,i}\sim N(\mu_{4,i},\sigma^2_{4,i})$ and $\hat{\mu}_{2,i}\sim N(\mu_{2,i},\sigma^2_{2,i})$.
Deriving the probability density function (PDF) of $\mathrm{\hat{Kur}}_i$ is challenging, as it involves the ratio of a Gaussian random variable to the square of another Gaussian variable. To the best of our knowledge, no closed-form expression for this PDF exists in the literature. To proceed with the BEP analysis for the Kurtosis-based detector, we reformulate expression (\ref{eq: BEPK}) as follows
\begin{align}
\label{eq: BEPK1}
&\mathrm{p}_{b,2}=\frac{1}{2}\mathrm{p}\{\hat{\mu}_{4,1}>\alpha(\hat{\mu}_{2,1})^2\}+\frac{1}{2}\mathrm{p}\{\hat{\mu}_{4,2}>\alpha(\hat{\mu}_{2,2})^2\},
\end{align}
where $\alpha\triangleq\mathrm{Th}_k$. Both terms in (\ref{eq: BEPK1}) are in the form of $\mathrm{p}\{Y>\alpha X^2\}$, where $X,Y\sim N(\mu_x,\sigma^2_x), N(\mu_y,\sigma^2_y)$. For simplicity of calculations, we assume $Y=\hat{\mu}_{4,i}$ is independent of $X=\hat{\mu}_{2,i}$. The probabilities are equal to
\begin{align}
\label{eq: pprob}
&\mathrm{P}\{Y>\alpha X^2\}=\int_{-\infty}^{+\infty}\int_{\alpha x^2}^{+\infty}f_X(x)f_Y(y)dydx\nonumber\\
&=\int_{=\infty}^{+\infty}f_X(x)(\int_{\alpha x^2}^{+\infty}f_Y(y)dy)dx\nonumber\\
&=\int_{=\infty}^{+\infty}f_X(x)\mathrm{Q}\Big(\frac{\alpha x^2-\mu_y}{\sigma_y}\Big)dx.
\end{align}

To further proceed with the calculations, we use the approximation of the Q-function as $\mathrm{Q}(x)\approx\exp(-ax^2-bx-c)$ in (\ref{eq: pprob}). 
\begin{align}
\label{eq: pprob1}
&\mathrm{P}\{Y>\alpha X^2\}=\frac{1}{\sigma_x\sqrt{2\pi}}\int_{-\infty}^{+\infty}\exp(q_4(x))dx,
\end{align}
where we have
\begin{align}
q_4(x)=Ex^4+Fx^2+Gx+H,
\end{align}
where $E\triangleq \frac{-a\alpha^2}{\sigma^2_y}$, $F\triangleq \frac{-1}{2\sigma^2_x}+\frac{2a\alpha\mu_y}{\sigma^2_y}-\frac{b\alpha}{\sigma_y}$, $G\triangleq \frac{\mu_x}{\sigma^2_x}$, and $H\triangleq \frac{-\mu^2_x}{2\sigma^2_x}-\frac{a\mu^2_y}{\sigma^2_y}+\frac{b\mu_y}{\sigma_y}-c$. Using (\ref{eq: pprob1}) into (\ref{eq: BEPK1}), the BEP of the Kurtosis-based detector is equal to
\begin{align}
\mathrm{p}_{b,2}&=\frac{1}{2\sigma_{x,0}\sqrt{2\pi}}\int_{-\infty}^{+\infty}\exp(q_{4,0}(x))dx+\nonumber\\
&\frac{1}{2\sigma_{x,1}\sqrt{2\pi}}\int_{-\infty}^{+\infty}\exp(q_{4,1}(x))dx,
\end{align}
where $\sigma_{x,0}=\sigma_{\hat{\mu}_{2,0}}=\sigma_{2,0}$, $\sigma_{x,1}=\sigma_{\hat{\mu}_{2,1}}=\sigma_{2,1}$, and we have
\begin{align}
q_{4,j}(x)=E_jx^4+F_jx^2+G_jx+H_j,\quad j=0,1
\end{align}
where $E_j=\frac{-a\alpha^2}{\hat{\mu}_{4,i}}$, $F_j=\frac{-1}{2\sigma^2_{2,j}}+\frac{2a\alpha\mu_{\hat{\mu}_{4,j}}}{\sigma^2_{\hat{\mu}_{4,j}}}-\frac{b\alpha}{\sigma_{\hat{\mu}_{4,j}}}$, $G_j=\frac{\mu_{\hat{\mu}_{2,j}}}{\sigma^2_{\hat{\mu}_{2,j}}}$, and $H_j=\frac{-\mu^2_{\hat{\mu}_{2,j}}}{2\sigma^2_{\hat{\mu}_{2,j}}}-\frac{a\mu^2_{\hat{\mu}_{4,j}}}{\sigma^2_{\hat{\mu}_{4,j}}}+\frac{b\mu_{\hat{\mu}_{4,j}}}{\sigma_{\hat{\mu}_{4,j}}}-c$.

To calculate the optimum threshold for the Kurtosis detector, we should minimize the BEP. By enforcing the derivative of the BEP, we can find the optimum threshold $\alpha_{opt}$. Hence, taking the derivative of (\ref{eq: BEPK1}) and using (\ref{eq: pprob1}) and enforcing the derivative to be zero, we have
\begin{align}
\label{eq: alphaprob}
&\frac{1}{\sigma_{x,1}}\exp(E_1(\alpha)x_1^4+F_1(\alpha)x_1^2+G_1x_1+H_1)+\nonumber\\
&\frac{1}{\sigma_{x,2}}\exp(E_2(\alpha)x_2^4+F_2(\alpha)x_2^2+G_2x_2+H_2)=0,
\end{align}
where $x_j=\hat{\mu}_{2,j}$, and we have $E_j=E_j(\alpha)$, $F_{j}=F_{j}(\alpha)$, and the other parameters are not related to $\alpha$. The optimum $\alpha$ is obtained by solving the nonlinear equation in (\ref{eq: alphaprob}). The details are omitted for the sake of simplicity. In brief, it results in a quadratic equation in terms of $\alpha$ in the form of $c_1\alpha^2+c_2\alpha+c_1=0$. The positive root of this equation is the optimum threshold $\alpha_{opt}$.

\section{Simulation Results}
\label{sec: Simulation}

In this section, the performance of the proposed 8-ary 3D noise modulation scheme is investigated. In all schemes, we use $m_L=-0.1=-\mu$ and $m_H=+0.1=\mu$ with $\mu=0.1$. The parameters of Gaussian mixture models are $\sigma_{00}=\sigma_{0L}=2e-3$, $\sigma_{10}=\sigma_{1L}=50\sigma_{00}=1e-1$, $\sigma_{01}=\sigma_{0H}=2\sigma_{00}=4e-3$. The fourth parameter $\sigma_{11}=\sigma_{1H}=50\sigma_{01}=2e-1$. The probability parameters are equal to $p_L=0$ and $p_H=0.5$. For a fair comparison, we set the average transmit power to be equal for all three schemes, ensuring they are compared at the same signal-to-noise ratio (SNR).
The thresholds are set as defined in Section \ref{sec: MoG}. Unless otherwise stated, the noise standard deviation is $\sigma_w=5\times10^{-5}$ and the number of samples per bit duration is $N=60$. The total number of symbols (or number of bits in each sub-channel) in the simulations is assumed to be $K=20000$, which is averaged over 20 Monte Carlo simulations. The performance is evaluated using the general and individual BEPs: $\mathrm{p}_{b,0}$, $\mathrm{p}_{b,1}$, and $\mathrm{p}_{b,2}$.

\begin{figure}
\begin{center}
\includegraphics[width=7cm]{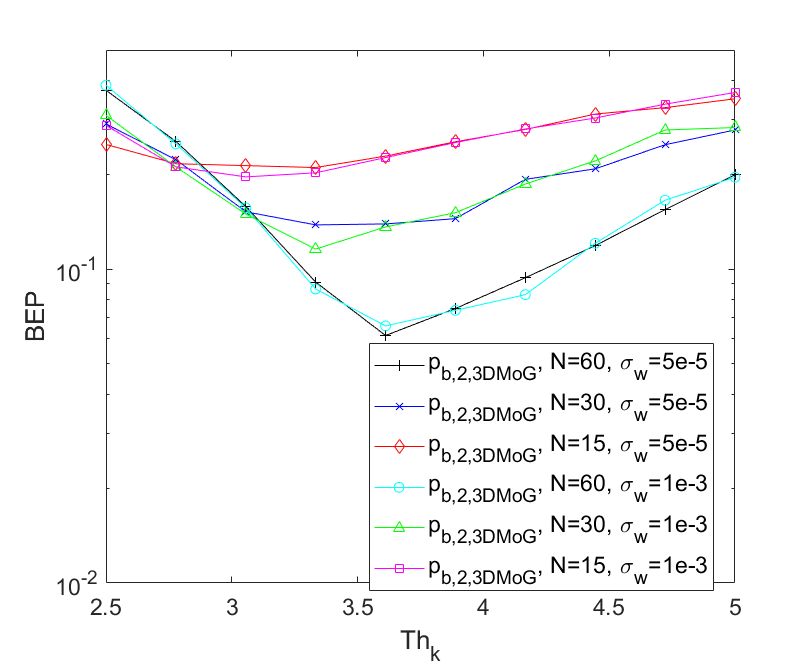}
\end{center}
\vspace{-0.5 cm}
\caption{Simulated BEPs of the proposed 3D-8ary MoGNM Modulator versus the threshold of Kurtosis $\mathrm{Th}_k$ in two cases of $\sigma_w=5e-5$ and $\sigma_w=1e-3$, and three cases of $N=15$, $N=30$, and $N=60$.}
\label{fig4}
\vspace{-0.5 cm}
\end{figure}

Three experiments were conducted. The first experiment, shown in Fig.~\ref{fig4}, investigates the impact of the kurtosis threshold $\mathrm{Th}_K$ under different noise conditions and sample sizes. The threshold was varied between 2.5 and 5, with two noise standard deviations ($\sigma_w = 5 \times 10^{-5}$ and $\sigma_w = 1 \times 10^{-3}$) and three sample sizes ($N = 15$, $N = 30$, and $N = 60$). For fair comparison, the same sampling rate was maintained for both the 2D GQNM and 3D-8ary-MoG-NM schemes.

The results show optimal thresholds of approximately $\alpha = 3.1$, $\alpha = 3.3$, and $\alpha = 3.6$ for $N = 15$, $N = 30$, and $N = 60$, respectively. These experimental values show good agreement with the theoretical optimum thresholds of 3.23, 3.42, and 3.75 derived in subsection~\ref{sec: BEPKur}. The noise standard deviation was found to have a negligible impact on performance, while increasing the sample size $N$ consistently reduced the BEP.

\begin{figure}
\begin{center}
\includegraphics[width=7cm]{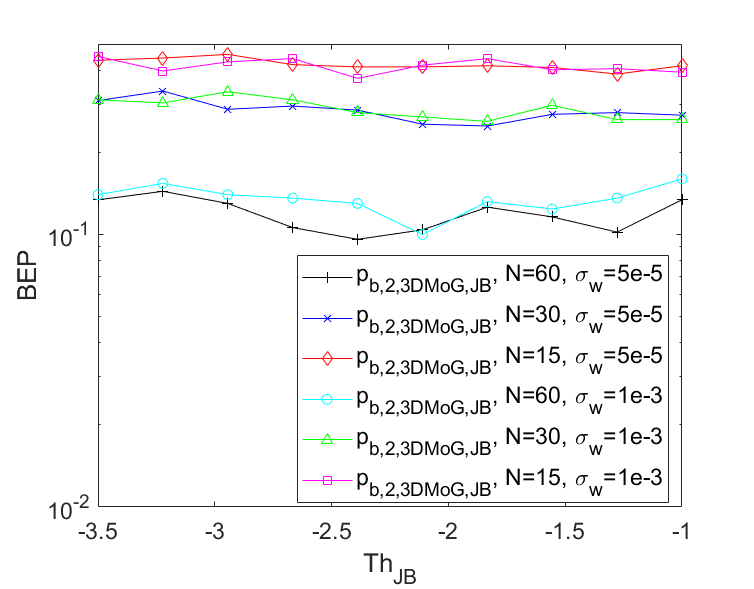}
\end{center}
\vspace{-0.5 cm}
\caption{Simulated BEPs of the proposed 3D-8ary MoGNM Modulator versus the threshold of JB $\mathrm{Th}_{JB}$ in two cases of $\sigma_w=5e-5$ and $\sigma_w=1e-3$, and three cases of $N=15$, $N=30$, and $N=60$.}
\label{fig41}
\vspace{-0.5 cm}
\end{figure}

We also implemented a detector based on JB statistics, which did not outperform the kurtosis-based approach. The relationship between the JB threshold $\mathrm{Th}_{JB}$ and BEP is shown in Fig.~\ref{fig41}, indicating that the JB threshold has less influence on BEP compared to the kurtosis threshold. The optimal JB threshold appears to lie between -2.4 and -2.0; for subsequent experiments, we used $\mathrm{Th}_{JB} = -2.4$.

%

\begin{figure}
\begin{center}
\includegraphics[width=7cm]{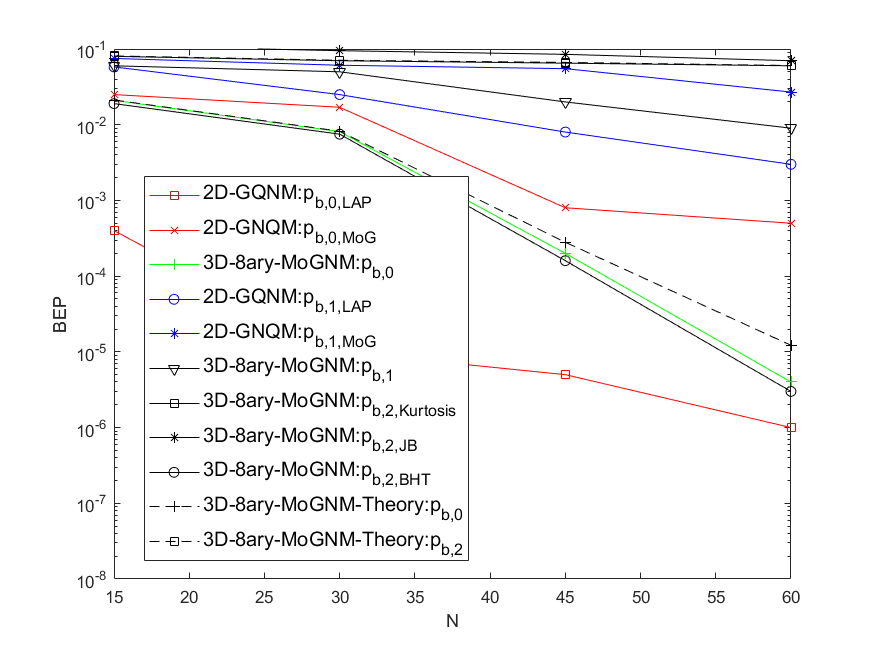}
\end{center}
\vspace{-0.5 cm}
\caption{Simulated BEPs of the proposed 3D-8ary MoGNM and 2D-GQNM modulators versus $N$.}
\label{fig5}
\vspace{-0.5 cm}
\end{figure}

The second experiment, illustrated in Fig.~\ref{fig5}, compares simulated and theoretical BEPs (shown with dashed lines) as functions of the sample size $N$, using fixed thresholds $\mathrm{Th}_k = 3.5$ and $\mathrm{Th}_{JB} = -2.5$. This analysis demonstrates the effect of $N$ on system performance across the evaluated range $15 \leq N \leq 60$. Consistent with previous findings, increasing $N$ reduces BEP across all schemes.

Notably, the 2D Laplacian scheme achieves the lowest BEP in the first sub-channel ($\mathrm{p}_{b,0}$), while the BHT detector in the 3D-8ary MoGNM scheme yields the second-best performance for $\mathrm{p}_{b,2}$. The second sub-bit channel ($\mathrm{p}_{b,1}$) maintains acceptably low error rates. In contrast, the third sub-bit channel ($\mathrm{p}_{b,2}$), detected using either Kurtosis or JB statistics, exhibits the highest BEP. This performance limitation stems from the inherent difficulty of distinguishing non-Gaussian from pure-Gaussian distributions using single higher-order statistics.

Although the Kurtosis-based detector demonstrates higher BEP than the BHT approach, it offers significantly lower computational complexity by avoiding the extensive logarithmic and exponential operations required by BHT. The results show good agreement between theoretical and simulated BEP curves, though a slight discrepancy emerges at $N = 60$ due to limitations in the Monte Carlo simulation sample size.

\begin{figure}
\begin{center}
\includegraphics[width=7cm]{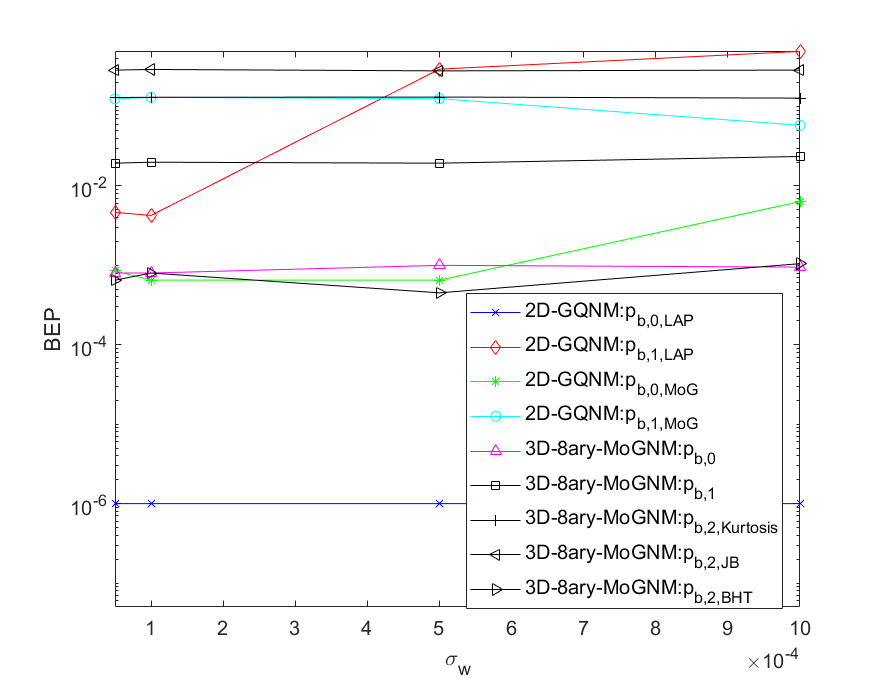}
\end{center}
\vspace{-0.5 cm}
\caption{Simulated BEPs of the proposed 3D-8ary MoGNM and 2D-GQNM modulators versus $\sigma_w$.}
\label{fig51}
\vspace{-0.5 cm}
\end{figure}

The third experiment, presented in Fig.~\ref{fig51}, compares the simulated BEP performance against channel noise variance $\sigma_w$ for a fixed sample size $N = 30$, using thresholds $\mathrm{Th}_k = 3.5$ and $\mathrm{Th}_{JB} = -2.4$. This analysis examines the impact of $\sigma_w$ across a range from $1 \times 10^{-6}$ to $1 \times 10^{-3}$, showing that BEP generally increases or remains stable as $\sigma_w$ grows, indicating that higher channel noise variance degrades detection performance.

Consistent with previous experiments, the 2D Laplacian scheme achieves the lowest BEP in the first sub-channel ($\mathrm{p}_{b,0,LAP}$), followed by the BHT detector in the 3D-8ary MoGNM scheme for $\mathrm{p}_{b,2}$, which ranks second. The second sub-bit channel ($\mathrm{p}_{b,1}$) maintains acceptably low error rates. In contrast, the third sub-bit channel ($\mathrm{p}_{b,2}$), detected using Kurtosis or JB statistics, exhibits comparatively higher BEP.


%

\section{Conclusion and future work}
\label{sec: con}

This paper proposes a 3D 8-ary modulation scheme designed to enhance the data rate of noise modulators. The three modulation dimensions consist of the mean, variance, and mixture probability of a Gaussian distribution. The third dimension specifically controls the selection between a low-probability MoG distribution (approximating a pure Gaussian when $p_L = 0$) and a high-probability MoG distribution with $p_H$. This additional dimension enables the simultaneous detection of three sub-channels ($b_0$, $b_1$, and $b_2$) that modulate the mean, variance, and shape of the MoG distribution, respectively.

For detection, a threshold-based detector with closed-form BEP calculation is employed for $b_0$, while an ML detector is used for $b_1$. The detection of $b_2$ incorporates three different approaches: a thresholded Kurtosis-based detector, a thresholded JB-based detector, and a BHT detector. The BEP for both the Kurtosis-based and BHT detectors is derived in a tractable, though not closed-form, manner. The optimal threshold for the Kurtosis-based detector is also determined tractably.

Simulation results demonstrate that the Kurtosis-based detector achieves a minimum BEP of $\mathrm{p}_{b,2} = 0.06$, reflecting its limited but computationally efficient performance. In contrast, the BHT detector attains a lower BEP comparable to that of other sub-channels and existing 2D GQNM schemes. This 3D scheme effectively enables the embedding of a third bit within two bits by leveraging MoG distributions instead of pure Gaussian ones, while maintaining sufficiently low BEP levels.


\section*{Acknowledgement}
This work has been partially funded by the xGMobile Project (XGM-AFCCT-2024-9-1-1) with resources from EMBRAPII/MCTI (Grant 052/2023 PPI IoT/Manufatura 4.0), by CNPq (302085/2025-4, 306199/2025-4), and FAPEMIG (APQ-03162-24).

\end{document}